# Engineering Interlayer Hybridization in Energy Space via Dipolar Overlayers

Bin Shao[1,2], Xiao Jiang[2], Jan Berges[3], Sheng Meng[4,5,6], Bing Huang[2,7]*

1. College of Electronic Information and Optical Engineering, Nankai University, Tianjin 300350, China
2. Beijing Computational Science Research Center, Beijing 100193, China
3. Institut für Theoretische Physik, Bremen Center for Computational Materials Science, and MAPEX Center for Materials and Processes, Universität Bremen, Bremen D-28359, Germany
4. Beijing National Laboratory for Condensed Matter Physics and Institute of Physics, Chinese Academy of Sciences, Beijing 100190, China
5. Songshan Lake Materials Laboratory, Dongguan, Guangdong 523808, China
6. School of Physical Sciences, University of Chinese Academy of Sciences, Beijing 100049, China
7. Department of Physics, Beijing Normal University, Beijing 100875, China

E-mail: bing.huang@csrc.ac.cn

**The interlayer hybridization (IH) of van der Waals (vdW) materials is thought to be mostly associated with the unignorable interlayer overlaps of wavefunctions ($t$) in real space. Here, we develop a more fundamental understanding of IH by introducing a new physical quantity, the IH admixture ratio $\alpha$. Consequently, an exotic strategy of IH engineering in energy space can be proposed, i.e., instead of changing $t$ as commonly used, $\alpha$ can be effectively tuned in energy space by changing the onsite energy difference ($2\Delta$) between neighboring-layer states. In practice, this is feasible via reshaping the electrostatic potential of the surface by deposing a dipolar overlayer, e.g., crystalline ice. Our first-principles calculations unveil that IH engineering via adjusting $2\Delta$ can greatly tune interlayer optical transitions in transition-metal dichalcogenide bilayers, switch different types of Dirac surface states in $Bi_2Se_3$ thin films, and control magnetic phase transition of charge density waves in $1H/1T$-$TaS_2$ bilayers, opening new opportunities to govern the fundamental optoelectronic, topological, and magnetic properties of vdW systems beyond the traditional interlayer-distance or twisting engineering.**

The interlayer hybridization (IH) of stacked two-dimensional (2D) van der Waals (vdW) materials manifests itself as a unique tuning knob for their overall physical propertie[1–6]. Compared with the conventional bulk materials, this hybridization arises from the relatively weak vdW interaction, whose strength is comparable to that of external stimuli. Therefore, (i) it can be lifted by mechanical exfoliation[7], opening the door for atomically thin 2D materials; (ii) it can also be utilized to manipulate the electronic properties of vdW homo/hetero-structures, not only to engineer their band structures[3,8], but also to drive a variety of many-body phenomena such as magnetic phase transitions[9,10], superconductivity[11], interlayer excitons[5,12,13], and charge density waves (CDWs)[14,15]; (iii) it can even be applied to realize twist engineering, bringing the studies of superconducting[16] or strongly correlated phases[17] in the 2D limit to a new level. Thus, searching an efficient way of tuning the IH always plays a central role for both fundamental studies and practical applications of stacked vdW materials.

In general, the electronic states of each monolayer component in vdW-stacked systems can be considered as states confined in a vertical finite-depth quantum well. As demonstrated in Fig. 1a (upper panel), the states in each well cannot be fully confined, leading to unignorable overlap of their wavefunctions in real space, referred to as $t$, between the quantum wells of different layers. Usually, the IH is tacitly accepted to be mostly associated with $t$, whose value can be effectively tuned in real space by mechanical exfoliation[7], stacking modulation[14,15], and twisting[16,17], resulting in scientific (i)-(iii) discoveries discussed above. A fundamental question is raised here: Besides tuning $t$ as commonly used, is there an alternative way to tune the IH? The answer to this question relies on a more fundamental understanding of the IH.

In addition to $t$ in real space, as shown in the Fig. 1a (bottom panel), there can be an offset energy, referring as $2\Delta$, in the onsite energy of neighboring-layer states in energy space. Similar to real-space $t$, this $2\Delta$ can be understood as the energy-space "overlap" of the coupled states. Considering these two overlaps and summarizing all the $k$-dependence of the initial energy dispersion in $\varepsilon_0(k)$, the IH can be illustrated in a minimal two-band model involving states $|a\rangle = (1,0)$ and $|b\rangle = (0,1)$ belonging to the A and B layer, respectively. The resulting Hamiltonian reads $\widehat{H}(k) = \varepsilon_0(k)\mathbf{1} + t\sigma_1 - \Delta\sigma_3$, where the $\sigma_i$ refer to the Pauli matrices. The eigenstates of the Hamiltonian are $|+\rangle = \cos(\varphi)|b\rangle + \sin(\varphi)|a\rangle$ and $|-\rangle = \cos(\varphi)|a\rangle - \sin(\varphi)|b\rangle$, respectively, with $\tan 2\varphi = t/\Delta$ and corresponding energies $\varepsilon_\pm(k) = \varepsilon_0(k) \mp \sqrt{\Delta^2 + t^2}$. The IH yields an admixture of the wavefunctions of $|a\rangle$ and $|b\rangle$. To describe the degree of this admixture, we define a key physical quantity, the IH admixture ratio $\alpha$, as the difference of the eigenstates' weights on $|a\rangle$ and $|b\rangle$, which reads $\alpha = 1 - \frac{|\langle\pm|a\rangle\langle a|\pm\rangle - \langle\pm|b\rangle\langle b|\pm\rangle|}{\langle\pm|a\rangle\langle a|\pm\rangle + \langle\pm|b\rangle\langle b|\pm\rangle} = 1 - 1/\sqrt{1 + 4(t/2\Delta)^2}$ with $t/2\Delta \in [0, +\infty]$. Importantly, the $\alpha$ unveils a more fundamental understanding of the IH, i.e., the IH is instead prone to the ratio of the two overlaps $t/2\Delta$, not merely related to $t$ as generally accepted.

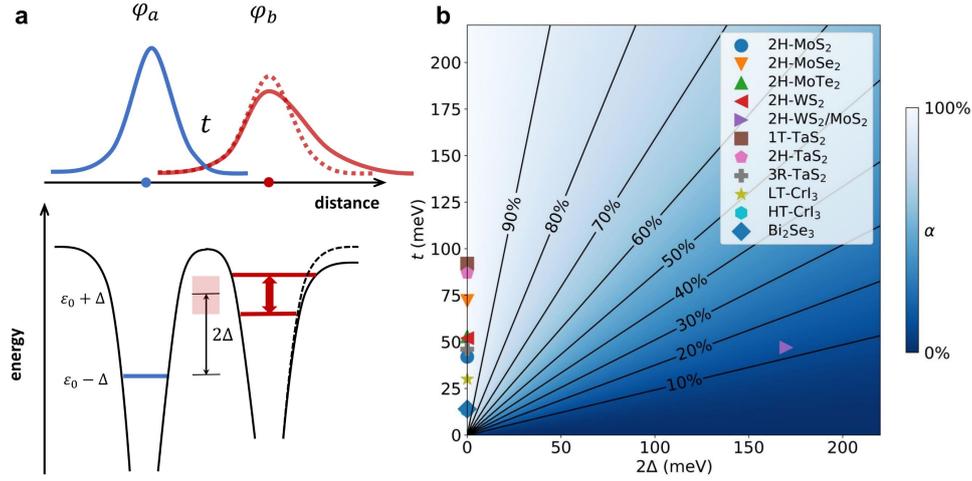

**Fig.1 IH engineering in stacked vdW systems. (a)** Schematic diagram of the interlayer hybridization (IH) in real space (upper panel) and energy space (bottom panel). Reshaping the electrostatic potential, denoted as dashed lines, leads to the variation of the onsite energy of layer B, along with the change of onsite energy difference $2\Delta$ between neighboring layers. **(b)** Phase diagram of the IH admixture ratio $\alpha$ as a function of $t$ and $2\Delta$, where the calculated $\alpha$ of several commonly used 2D systems are also indicated.

The phase diagram of $\alpha$ as a function of $t$ and $2\Delta$ is depicted in Fig. 1b, where the two parameters of several commonly used homo/hetero-structures are also shown (see full data in Supplemental Table 1). For homo-structures, $2\Delta = 0$, resulting in the highest admixture ratio $\alpha = 100\%$ of electronic states between neighboring layers. In contrast, the $2\Delta \neq 0$ in hetero-structures, thus, the corresponding $\alpha$ can be relatively small, e.g., $2\Delta = 200$ meV results in $\alpha = 12.5\%$ in 2H-MoS$_2$/WS$_2$. Interestingly, the phase diagram strongly indicates that even without changing $t$, the $\alpha$ of vdW systems can still be effectively weakened or enhanced by varying $2\Delta$. As shown in Fig. 1a, we propose that tuning $2\Delta$ is feasible via asymmetrically reshaping the electrostatic potential of the stacked 2D systems, e.g., via deposing a dipole overlayer on the top or bottom surface, providing an alternative opportunity for IH engineering in energy space.

Crystalline ice, exhibiting the hexagonal Ih phase (Ih-ice) under a normal pressure[18], is one of most common dipole layers existing in nature. Here, we propose to engineer $\alpha$ via modulating $2\Delta$ instead of $t$, which can be achieved by deposing an Ih-ice overlayers on the upper surface of 2D vdW systems. Using first-principles calculations, we demonstrate that IH engineering via tuning $2\Delta$ can realize greatly tunable interlayer optical transitions in transition-metal dichalcogenide (TMD) bilayers, switchable different types of Dirac surface states in Bi$_2$Se$_3$ thin films, and controllable magnetic phase transition of CDW in 1H/1T-TaS$_2$ hetero-bilayers, which potentially open a new door to engineer the optoelectronic, topological, and magnetic properties of 2D vdW systems.

**Case I: Tunable interlayer optical transitions in TMD bilayers**
Band-edge optical transitions (BOTs) play a central role in determining the fundamental optoelectronic properties of semiconductors. In general, there are two types of BOTs in layered materials, i.e., interlayer BOTs and intralayer BOTs. In practice, it is interesting to realize a highly tunable interlayer BOT, that is fundamentally related to the magnitude of interlayer optical transition probabilities ($r^2$, squares of transition dipole moments) between the band-edge states, for controllable interlayer exciton and lighting[6,19]. Usually, the interlayer $r^2$ are the intrinsic properties of a system, which strongly depend on $\alpha$. Taking TMD bilayers as examples, as shown in Fig. 2a, the smaller (larger) $\alpha$ in the hetero-bilayer (homo-bilayer) indicates a weaker (stronger) IH admixture between different layers, which may consequently result in a weaker (stronger) interlayer $r^2$. Here, we propose that via $2\Delta$ engineering – tuning IH by changing $2\Delta$ – we may greatly tune $r^2(\alpha)$ in TMD bilayers.

To verify this idea, we construct a supercell with a $\sqrt{3} \times \sqrt{3}$ Ih-ice overlayer on a $\sqrt{7} \times \sqrt{7}$ TMD bilayer (see Supplemental Figs. 1 and 2) to minimize the lattice mismatch between them. First, we consider the hetero-bilayer $WS_2/MoS_2$. The calculated $2\Delta = 169$ meV gives rise to a small $\alpha = 12.5\%$ in $WS_2/MoS_2$. Fig. 2b shows the calculated unfolded orbital-resolved effective band structure of AB-stacked $WS_2/MoS_2$. Overall, the band-edge states in $WS_2/MoS_2$ exhibit a type-II band alignment. Interestingly, depositing a double-layer (2L) Ih-ice on top of $WS_2/MoS_2$ (Fig. 2c) effectively reduces $2\Delta$ from 169 to 57 meV, leading to a large enhancement of $\alpha$ from 12.5% to 47.9%. More importantly, although the band structures before and after Ih-ice deposition are very similar, the interlayer $r^2$ is significantly enhanced (Fig. 2g) with respect to the ice-free (0L) case (Fig. 2f), consistent with the mechanism proposed in Fig. 2a.

Second, we consider the homo-bilayer $MoS_2/MoS_2$. Since $2\Delta = 0$ in $MoS_2/MoS_2$ (Fig. 2d), $\alpha = 100\%$, leading to the electronic states from different $MoS_2$ layers being energetically degenerated. The $\alpha = 100\%$ in $MoS_2/MoS_2$ induces large band splittings around the $\Gamma$ point in the top of the valence band (VB) and along the paths $\Gamma$-$K$/$\Gamma$-$M$ in the bottom of the conduction band (CB), which are absent in monolayer $MoS_2$ (see Supplemental Fig. 3). The situation is dramatically changed when a 2L Ih-ice is deposited on top of $MoS_2/MoS_2$. As shown in Fig. 2e, the bands from different layers are rigidly shifted with respect to each other, forming an ideal type-II band alignment. The calculated $2\Delta = 222$ meV gives rise to a dramatic reduction of $\alpha$ from 100% to 7.6%. This rather small $\alpha$ can not only effectively eliminate the original band splitting in the ice-free (0L) case (Fig. 2d), but can also convert the band structure of the top $MoS_2$ layer in $MoS_2/MoS_2$ from indirect bandgap to direct bandgap, similar to that of monolayer $MoS_2$. More importantly, compared with the 0L case (Fig. 2h), the interlayer $r^2$ in the 2L case is greatly suppressed to almost zero due to the largely reduced $\alpha$ (Fig. 2i), consistent with the mechanism proposed in Fig. 2a. Therefore, via $2\Delta$ engineering, we may strongly enhance (suppress) the interlayer BOTs for hetero-bilayer (homo-bilayer) TMDs, yielding the corresponding enhancement (suppression) in its low energy adsorption spectroscopy (see Supplemental Fig. 4).

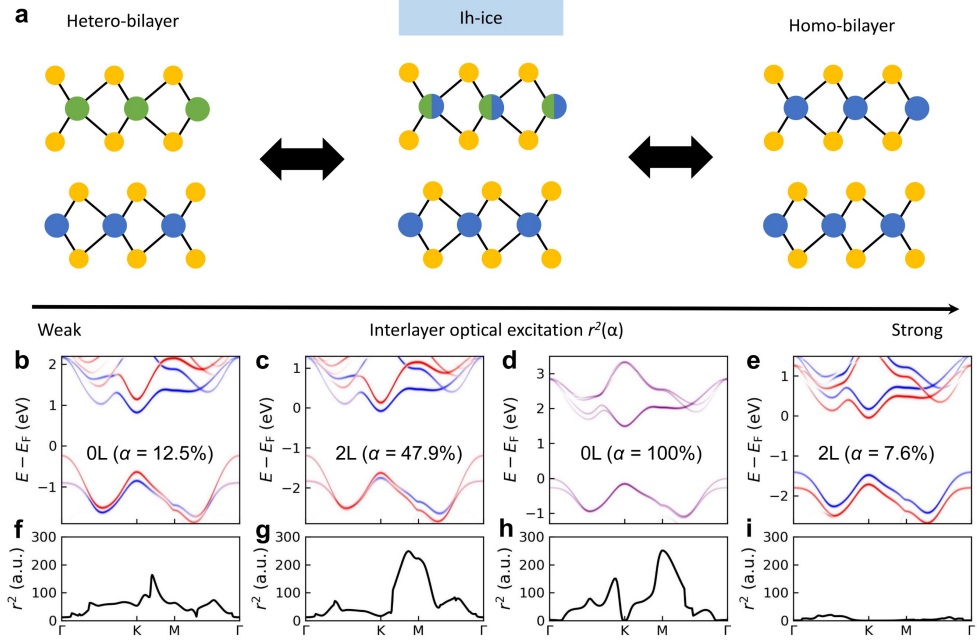

**Fig.2 Optoelectronic engineering in TMD bilayers. (a)** Schematic diagram of tuning interlayer optical transition probabilities ($r^2$) in TMD bilayers. Interlayer $r^2$ in TMD hetero- and homo-bilayers are weak and strong, respectively. Via $2\Delta$ engineering, the interlayer $r^2(\alpha)$ of hetero- and homo-bilayers can be greatly enhanced and weakened, respectively. Orbital-resolved effective band structures of $WS_2/MoS_2$ covered by **(b)** 0L and **(c)** 2L Ih-ice. Electronic states of top $WS_2$ and bottom $MoS_2$ layers are highlighted by red and blue colors, respectively. **(d)-(e)** Same as **(b)-(c)** but for the cases of $MoS_2/MoS_2$. Electronic states of top and bottom $MoS_2$ layers are highlighted by blue and red colors, respectively. **(f)-(g)** Calculated interlayer $r^2$ between four band-edge states in the band structures of $WS_2/MoS_2$ shown in **(b)-(c)**, respectively. **(h)-(i)** Same as **(f)-(g)** but for the case of $MoS_2/MoS_2$.

**Case II: Switchable different types of Dirac fermions in $Bi_2Se_3$ thin films**

Topological materials, starting from topological insulators (TIs), are standing in the cutting edge of quantum materials[20–22]. An ideal TI, e.g., $Bi_2Se_3$, is a quantum matter with a bulk gap and an odd number of relativistic Dirac fermions on the surface, in which the bulk is insulating but the surface can conduct electric current with helical spin texture[23,24]. However, it is known that Dirac surface states can only be realized in a thick $Bi_2Se_3$, i.e., more than six quintuple-layer (QL). Once the thickness of $Bi_2Se_3$ is less than six QL, the IH between the top and bottom Dirac surface states can induce a hybridization gap (Fig. 3a, left and middle panels), destroying the massless feature of Dirac surface states[25,26]. As shown in Fig. 1b, via $2\Delta$ engineering, we may reduce the $\alpha$ in $Bi_2Se_3$ ultrathin films, which may lift the coupling between top and bottom surface states (Fig. 3a, right-panel) and result in a transition from massive to massless Dirac fermions.

To verify this idea, as shown in Fig. 3b, we have selected 3QL-$Bi_2Se_3$ ultrathin films as an example, in which the top of 3QL-$Bi_2Se_3$ is covered by an Ih-ice overlayer. In the ice-free case (0L), $2\Delta = 0$, therefore, $\alpha = 100\%$. The hybridization between the top and bottom Dirac surface states in 3QL-$Bi_2Se_3$ can induce a large hybridization gap of about 50 meV between them, transforming the original massless Dirac fermions to massive ones[25]. Moreover, because of the effective surface-state coupling, the spin texture of massive Dirac surface states, which are completely degenerate in energy space, is far away from the ideal helical one (inset of Fig. 3c), i.e., the directions of the spin moments are not perpendicular to that of the crystal momenta.

When a 2L Ih-ice overlayer is deposited on the top of 3QL-$Bi_2Se_3$ slab (see Supplemental Fig. 5), it can result in $2\Delta = 239$ meV, which can consequently reduce $\alpha$ from 100% to almost 0%. As a result, the surface states from top and bottom layers are no longer degenerate in energy space. And the Dirac fermions on the top and bottom surfaces are restored to be massless. Interestingly, the spin textures of the Dirac surface states, which are fully separated in the energy space, are also restored to the ideally helical one, i.e., the directions of the spin moments are rigorously perpendicular to that of the crystal momenta. Therefore, an interesting transition from massive to massless Dirac fermions can be well achieved by $2\Delta$ engineering, along with the modulation of its spin texture.

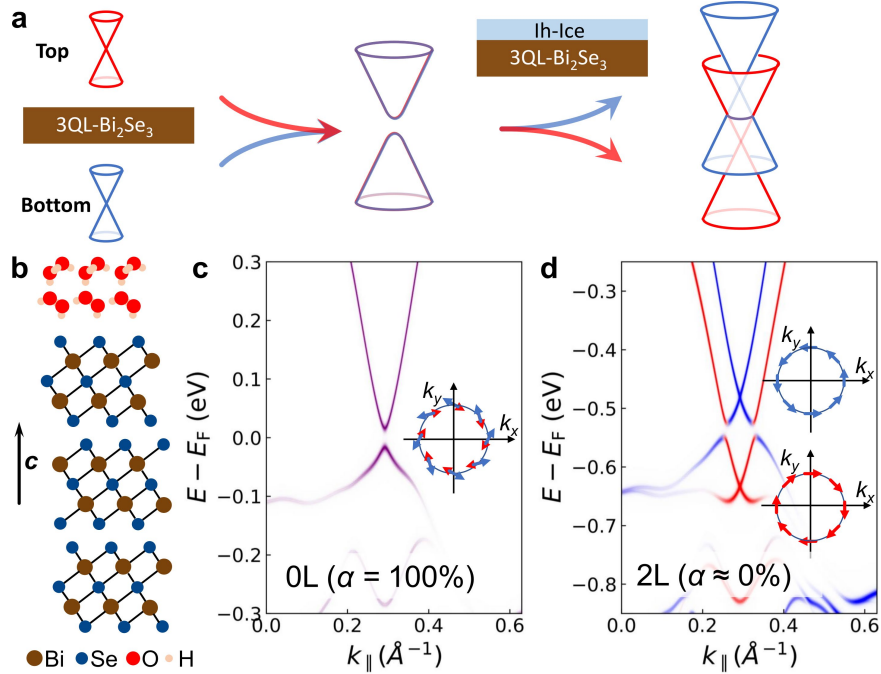

**Fig. 3 Dirac surface states engineering in $Bi_2Se_3$ thin films. (a)** Schematic diagram of switchable massive-massless Dirac fermions in 3QL-$Bi_2Se_3$. Because of the strong orbital hybridization, the surface states of 3QL-$Bi_2Se_3$ open a gap and Dirac fermions are massive (left panel to middle panel). By deposing the Ih-ice overlayer, this hybridization is expected to be effectively reduced, thus, the surface Dirac fermions will be recovered to be massless (middle panel to right panel). **(b)** Side view of atomic structure of 3QL-$Bi_2Se_3$ covered by 2L Ih-ice. Orbital-resolved effective band structure of 3QL-$Bi_2Se_3$ with **(c)** 0L and **(d)** 2L Ih-ice overlayer. Weights of the surface states from the top and bottom sides are highlighted by red and blue colors, respectively. Insets: spin textures of the surface states from top (red arrow) and bottom (blue arrow) sides. Shift of the surface states to lower energies in **(d)** results from the charge transfer from the Ih-ice overlayer to the $Bi_2Se_3$ (see Supplemental Fig. 6).

**Case III: Controllable magnetic phase transition of CDW in 1T/1H-TaS$_2$**

Magnetism is one of the central phenomena in condensed matter physics, which is mainly determined by the spin exchange interaction $J$ in the system. The realization of controllable $J$ is the key to realize controllable magnetic phase transition. A 1T/1H-TaS$_2$ heterostructure has recently been proposed to be a promising platform to simulate a Kondo lattice[27,28]. The 1T-TaS$_2$ in the Star-of-David (SoD) charge-density wave (CDW) phase naturally forms a large unit cell with a single localized spin moment. These localized spin moments can interact with each other through the $5d$ conduction electrons in metallic 1H-TaS$_2$ via the IH. Thus, the $J$ between localized spin moments and conduction electrons highly depends on $\alpha$, i.e., $J$ is a function of $\alpha$ (Fig. 4a). By $2\Delta$ engineering, we may effectively control $J(\alpha)$ in 1T/1H-TaS$_2$, which can consequently control the magnetic phase transition of CDW in 1T/1H-TaS$_2$.

To verify this idea and to model the 1T/1H-TaS$_2$ heterostructure with the SoD-CDW phase, we combine a $\sqrt{39} \times \sqrt{39}$ 1T-TaS$_2$ structure with a $6 \times 6$ 1H-TaS$_2$ structure (see Supplemental Fig. 7). The calculated local spin moment on the Ta atom in the center each SoD cluster is about 0.2 $\mu_B$, and its direction can be flipped freely (there is an energy barrier of about 0.02 meV per SoD cluster), which results in a paramagnetic phase. As shown in Fig. 4b, the calculated density of states (DOS) indicates that the half-occupied state contributed by the Ta-$d_{z^2}$ in the center of SoD in 1T-TaS$_2$ is pushed upwards to higher energies, while the states derived from the Ta-$d$ in 1H-TaS$_2$ are pushed downwards to lower energies. This noticeable level repulsion between 1T and 1H layers around the Fermi level manifests the existence of an unignorable IH, as also indicated by the calculated $\alpha = 13.9\%$. Considering both the spin-polarization and DOS calculations, it can be concluded that the 1T/1H-TaS$_2$ heterostructure probably falls into the Kondo screened regime in the phase diagram in Fig.4c, consistent with the experimental observations[28].

If 2L Ih-ice is deposited on top of the 1T-TaS$_2$ layer, the increased $2\Delta$ can weaken $\alpha$. Interestingly, as shown in Fig. 4b, although $\alpha$ is slightly reduced from 13.9% to 10.2%, it can strongly suppress the level repulsion between the 1T- and 1H-TaS$_2$ layer. Instead, the Kondo spin screening of the local moments will be dominated by the RKKY interaction[29], leading to a magnetic order. Indeed, with the assistance of Ih-ice overlayer, the freely flipped localized spin moment in the 1T-TaS$_2$ layer becomes magnetically ordered with the local magnetic moment of the center Ta in each SoD cluster being of about 0.3 $\mu_B$ (left in Fig. 4c). During the self-consistent iterations, all the different initial magnetic configurations will eventually converge to a collective order (left in Fig.4c), indicating a (strong) local energy minimum. Thus, a novel magnetic phase transition of CDW in the 1T/1H-TaS$_2$ heterostructure can be well controlled by $2\Delta$ engineering.

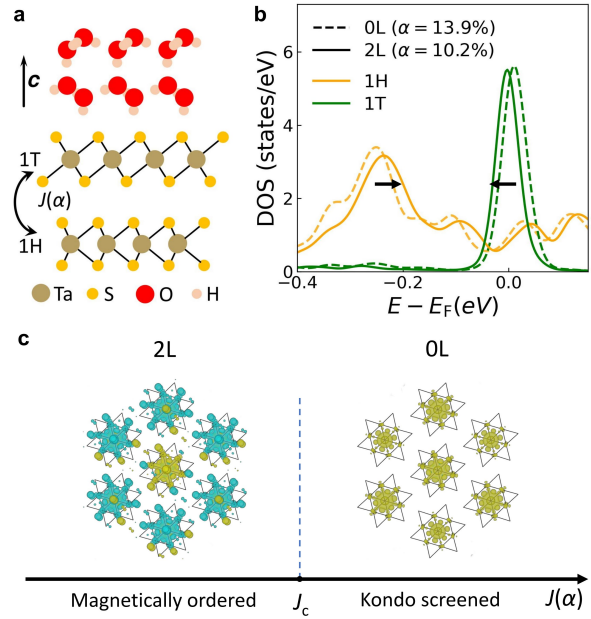

**Fig. 4 Engineering magnetic phase transition in 1T/1H-TaS$_2$ heterostructure. (a)** Side view of atomic structure of 1T/1H-TaS$_2$, where the top of the 1T layer is covered by Ih-ice. $J$ is a function of $\alpha$. **(b).** Projected DOS of 1T/1H-TaS$_2$ with 0L (dashed line) and 2L (solid line) Ih-ice overlayer. The projection on $d_{z^2}$ orbital of the central Ta in the SoD of the 1T-TaS$_2$ layer is highlighted by green color, while the projection on the $d_{z^2}$, $d_{xy}$, and $d_{x^2-y^2}$ orbitals of Ta in the 1H layer is highlighted by orange color. **c.** Schematic diagram of the magnetic phase transition between a magnetically ordered phase induced by RKKY interaction and a paramagnetic phase resulting from the Kondo screening. Left and right panel shows the calculated spin density maps of the 1T/1H-TaS$_2$ heterostructure with 2L and 0L Ih-ice overlayer, respectively.

**Discussion**

The critical factor of the IH engineering in energy space proposed here is to find a feasible approach to strongly reshape the local electrostatic potential of a surface layer. In principle, candidates for the dipole overlayers can generally be any materials with a net dipole moment perpendicular to the surface. The choice of an Ih-ice overlayer in this work is based on two concerns. First, the water molecules possess a large dipole moment (1.8-3.0 Debye)[30], being capable of sharply tailoring the surface electrostatic potential. Second, water molecules can be frozen into a crystalline phase with its polarization controlled by external fields[31–33]. Although there are reports on polar molecules as adsorbates influencing the electronic structures of solid surfaces[34–37], the adsorbate-induced variations in the IH admixture ratio is barely explored and poorly understood.

In summary, beyond the common belief that IH is mostly associated with the wavefunction overlap in real space $t$, we propose that IH can also be dramatically engineered in energy space by tuning the energy separation $2\Delta$. Via the proposed $2\Delta$ engineering, we demonstrate that the optoelectronic, topological, and magnetic properties of different 2D vdW homo/hetero-junctions can be well modulated. Our concept and material demonstrations may open the way for new ideas for the IH engineering in vdW systems beyond the conventional ways.

## Methods

### Computational details

For the density-functional theory (DFT) calculations, we use the Vienna Ab Initio Simulation Package (VASP)[38] with the projector-augmented-wave basis sets[39,40] and the generalized gradient approximation (GGA) to the exchange-correlation potential[41]. While the lattice constants of 1H-MoS$_2$, 1H-WS$_2$, 1T-TaS$_2$, and 1H-TaS$_2$ are fully optimized, the crystal structure of Bi$_2$Se$_3$ is taken from Ref. 42. To obtain a commensurate structure, we construct slabs of a $\sqrt{3} \times \sqrt{3}$ supercell of Ih-ice overlayer on a $\sqrt{7} \times \sqrt{7}$ supercell of MoS$_2$/MoS$_2$ (MoS$_2$/WS$_2$), a $1 \times 1$ Ih-ice overlayer on a $1 \times 1$ 3QL-Bi$_2$Se$_3$, a $\sqrt{21} \times \sqrt{21}$ supercell of Ih-ice overlayer on a $\sqrt{39} \times \sqrt{39}$ supercell of bilayer 1T-TaS$_2$ on a $6 \times 6$ supercell of 1H-TaS$_2$, where the Ih-ice overlayers are laterally compressed or expanded to eliminate the lattice mismatch. The in-plane lattice constant of the 1T/1H-TaS$_2$ supercell is set to be the average value of the $\sqrt{39} \times \sqrt{39}$ supercell of 1T-TaS$_2$ and the $6 \times 6$ 1H-TaS$_2$. A vacuum separation of more than 15 Å is used in all the slabs. In all the cases, the structures are relaxed until the forces acting on them are below 0.01 eV Å$^{-1}$. In the evaluation of the electronic structures, we choose a plane-wave cut-off energy of 400 eV for all the cases and a $\Gamma$-centered $7 \times 7 \times 1$, $15 \times 15 \times 1$, and $5 \times 5 \times 1$ $k$-mesh for the MoS$_2$/MoS$_2$ (MoS$_2$/WS$_2$), 3QL-Bi$_2$Se$_3$, and 1T/1H-TaS$_2$ case, respectively. The convergence criterion for the total energy is a change of less than $10^{-5}$ eV between self-consistency iterations. We have taken into account the vdW interaction via the D2 method of Grimme[43] in all the calculations, and we have applied a dipole correction[44] in the direction perpendicular to the slab in the cases with Ih-ice overlayer. The spin-orbit coupling is included in the calculation of the surface state and the spin texture of 3QL-Bi$_2$Se$_3$. The $+U$ correction[45] with $U - J = 2.27$ eV[46,47] has been applied to the Ta-$d$ orbital in the 1T layer.

### Effective Wannier Hamiltonian

We construct an effective Hamiltonian in the basis of atomic projected Wannier functions in the MoS$_2$/MoS$_2$, MoS$_2$/WS$_2$, and 3QL-Bi$_2$Se$_3$ cases using the Wannier90 package[48], see Supplemental Figs. 8, 9 and 10. We choose the W-$d$, Mo-$d$, and S-$p$ orbitals of MoS$_2$ and WS$_2$, O-$p$ orbitals of the Ih-ice, and Bi-$p$ and Se-$p$ orbitals of Bi$_2$Se$_3$ as initial projections in the disentanglement process. In the case of the TMD bilayers, the effective Hamiltonian of the supercell is further mapped to the TMD primitive cell using our developed routines in the elphmod package[49]. Here, the exact choice of the primitive cell does not have a noticeable influence on the resulting band structure, showing that the change of the onsite energy differences 2Δ due to the Ih-ice overlayer is homogeneous.

### Estimation of the admixture ratio $\alpha$

For the case of MoS$_2$/MoS$_2$, MoS$_2$/WS$_2$, and 3QL-Bi$_2$Se$_3$, we estimate the corresponding admixture ratio $\alpha$ from the expression $\alpha = 1 - 1/\sqrt{1 + 4(t/2\Delta)^2}$. We choose the parameter $t$ as the leading term of the nearest-neighbour hopping between the orbitals from different layers, which are $\{$Mo/W-$d_{z^2}^{\text{top}}$, Mo-$d_{z^2}^{\text{bottom}}\}$ and $\{$Bi-$p_z^{\text{top}}$, Bi-$p_z^{\text{bottom}}\}$ in the case of MoS$_2$/MoS$_2$ (WS$_2$/MoS$_2$) and 3QL-Bi$_2$Se$_3$, respectively. The parameter $2\Delta$ is obtained as the difference of the onsite energies between corresponding orbitals. Given the large number of atoms in the 1T/1H-TaS$_2$ system, we select an alternative way of estimating the admixture ratio $\alpha$ for computational feasibility. We integrate the density of states projected on the $d_{z^2}$ orbital of the Ta atom in the center of SoD cluster (Ta$^{\text{SoD}}$-$d_{z^2}$) and on the $d_{z^2}$ orbital of its nearest neighbor Ta atom in the 1H layer. The integration energy window is determined by the full width at half maximum of the Ta$^{\text{SoD}}$-$d_{z^2}$ peak.

The admixture ratio $\alpha$ is then calculated as $\alpha = 1 - |N^{1T} - N^{1H}|/(N^{1T} + N^{1H})$, where $N^{1T}$ and $N^{1H}$ are the integrated projected densities of states.

**Calculation of the optical transition dipole moments (TDMs)**

The effective Wannier Hamiltonians of the primitive cell of the TMD bilayers and their eigenfunctions are applied to calculate the band velocity $v_{nm}$ with $v_{nm} = \langle n, k|\partial_k H|m, k\rangle$ where $n, m$ are electronic band indices. To identify the interlayer contributions to the band velocity, we introduce the project operator by following $v_{gg'}^{\text{Inter}} = \langle g, k|\partial_k H|g', k\rangle = \sum_{n,m}\langle g, k|n, k\rangle\langle n, k|\partial_k H|m, k\rangle\langle m, k|g', k\rangle$, where the local orbitals $|g, k\rangle, |g', k\rangle$ belong to different layers. The TDMs $r_{nm}$ can be obtained with the relation $r_{nm} = v_{nm}/i\omega_{nm}$ ($m \neq n$), where $\omega_{nm}$ is the eigenvalue difference between $|n, k\rangle$ and $|m, k\rangle$. Here, the estimated $r^2$ (squares of TDMs) are the sum of four optical transition processes between the top two valence bands and the bottom two conduction bands. These calculations are performed in our homemade NOPSS package[50].

**Calculation of orbital-resolved effective band structures**

Using the effective Wannier Hamiltonian, we calculate the interpolated band structure $\epsilon_n(k)$ and its Bloch state $|n, k\rangle$ in the cases of MoS$_2$/MoS$_2$, MoS$_2$/WS$_2$, and 3QL-Bi$_2$Se$_3$. Then, we estimate the orbital-resolved effective band structure via the spectral function $A(\omega, k)$ on a localized orbital $|g, k\rangle$ defined as $A^g(\omega, k) = -2\text{Im}\frac{\langle n, k|g, k\rangle\langle g, k|n, k\rangle}{\omega - \epsilon_n(k) + i0^+}$, where $i0^+$ is an imaginary offset.

**Acknowledgements**

We acknowledge the support from the NSFC (No. 12088101), the NSAF (No. U1930402). J. B. acknowledges the financial support by the Deutsche Forschungsgemeinschaft (DFG) through EXC 2077. We acknowledge Drs. Wujun Shi, Fawei Zheng, and Tim Wehling for fruitful discussions. The calculations were performed at Tianhe2-JK at CSRC.